\documentclass{aa}
\usepackage{graphicx}
\usepackage{txfonts}
\begin{document}

   \title{INTEGRAL observations of the peculiar BeX System \\ \object{SAX J2103.5+4545}}

   \author{P. Blay \and P. Reig \and S. Mart\'\i nez N\' u\~ nez \and A. Camero \and P. Connell \and V. Reglero}
   \offprints{P. Blay}
   \institute{Institut de Ci\`encia de Materials, Universitat de
 Val\`encia, E--46071 Paterna-Valencia, Spain \\ \email{pere.blay@uv.es} }

\date{}
   
\abstract{

We present an INTEGRAL data analysis of the X-ray transient \object{SAX
J2103.5+4545} during two outbursts detected in December 2002.  The
INTEGRAL coordinates and error circle agree with the position of the
recently proposed optical counterpart. A power-law plus cut-off model
provided a good fit to the 4--150 keV spectrum yielding a photon index of
1.0$\pm$0.1, a cut-off energy $E_{\rm cut}=7.6\pm2.0$ keV and a folding
energy $E_{\rm fold}=30.9\pm2.5$ keV. The X-ray luminosity in the 4-150
keV energy range was found to be 6.0$\times 10^{36}$ erg s$^{-1}$,
assuming a distance of 6.5 kpc. This luminosity, together with the derived
photon index, indicate that the source is in a bright state.   A
354.9$\pm$0.5 second pulse period is measured. This value is significantly
smaller than previous measurements, indicating a long-term spin-up
episode. 

}

\maketitle{}

%
%________________________________________________________________

\section{Introduction}

\object{SAX J2103.5+4545} was discovered with BeppoSAX in 1997 when the source
underwent a bright outburst (Hulleman et al. \cite{hulleman98}). The
source was found to be an X-ray pulsar with a 358.61 second pulse period.
An absorbed power-law of index 1.27 was the best fit to the spectra. Two
more active phases are reported in the literature, in October 1999
and March 2001 (Baykal et al. 2000, 2002). Pulse arrival time analysis
allowed the determination of the orbital parameters of the system. The
orbital period was calculated to be 12.68$\pm$0.25 days and the
eccentricity 0.4$\pm$0.2 (Baykal et al. 2000). The distance to the source was
estimated to be $\sim$3 kpc. A bright state, with a
luminosity of the order of 10$^{36}$ erg s$^{-1}$ and strong orbital
modulation, and a faint state, with a luminosity of the order of 10$^{34}$
erg s$^{-1}$ and weak orbital modulation, were found (Baykal et al. 2002).
Because of its spectral properties and the presence of pulsations, this
source was tentatively classified as a high-mass X-ray binary system
(HMXRB). The presence of pulsations pointed to a neutron star as the
source of the X-ray emission, while its transient nature, showing orbital
modulation, suggested a possible BeX system. In a BeX system a strongly
magnetized neutron star orbits a Be star. The pulsed X-ray emission
results as a consequence of the misalignment of the magnetic axis and the
rotation axis and is produced by the accretion of matter, driven by the magnetic
field lines, onto the poles of
the compact companion of the system. The Be
star shows a dense envelope with a disk-like geometry, from which the
compact companion accretes matter. Recurrent and moderate ($L_x \approx 10^{35}-10^{36}$ erg s$^{-1}$)
increases of the X-ray flux modulated with the orbital period are known as
type I outbursts.

Recently, Reig et al. (\cite{reig04}) has reported the first optical and
infrared observations of this system leading to the identification of the
optical counterpart with a B0V star.

The INTErnational Gamma Ray Astrophysics Laboratory (INTEGRAL), is an ESA
mission dedicated to high-energy astrophysics. During November-December
2002, within the Performance and Verification phase (PV), the Cygnus X-1 region was
intensively monitored with the aim to check the performance of all the
instruments and subsystems on board INTEGRAL, and to perform the first
background studies. In this work we present an spectral, timing and
spatial analysis of the X-ray transient pulsar \object{SAX J2103.5+4545}
using observations from this period. Previous related work can be found in
Lutovinov et al. (\cite{luto03}) and Del Santo et al. (\cite{santo03}),
who showed preliminary results from the first detections of \object{SAX
J2103.5+4545} during  the PV phase and the first two Galactic Plane scans,
respectively. The capabilities of the INTEGRAL mission to perform a
detailed study of this kind of sources are shown.

\section{Observations and Data Analysis}

   There are three coded-mask
high-energy instruments on board INTEGRAL. The gamma-ray imager (IBIS), is
composed of two detector layers: ISGRI on top, which is sensitive to
$\gamma$-ray photons between 20 keV and 1 MeV (although the efficiency
drops to 50\% at 150 keV) and PICsIT, which is optimised to detect higher
energy photons (up to 10 MeV). IBIS yields an angular resolution of 12
arc-minutes (Ubertini et al. \cite{ibispaper}). SPI is a gamma-ray
spectrometer that operates in the energy range 20 keV to 8 MeV (Vedrene et
al. \cite{spipaper}). JEMX operates in the X-ray range 3--30 keV and it
is composed of two twin telescopes with a sensitive geometric area of
$\sim$ 500 cm$^2$ per unit (Lund et al. \cite{jemxpaper}). An Optical
Monitoring Camera (OMC) gives support to the high-energy instruments in
the optical V band (Mass-Hese et al. \cite{omcpaper}). All the instruments
are co-aligned, hence, strictly simultaneous multi-wavelength observations
of high-energy sources in a very wide energy range can be performed for the
first time. A detailed description of the mission can be found in Winkler
et al. (\cite{integral}).

From December 8  up to  December 29, 2002 (INTEGRAL revolutions 19 to
25), \object{SAX J2103.5+4545} was detected by all instruments on board INTEGRAL.
The total on-source time was approximately 500 ks for ISGRI, 730 ks for SPI
and 20 ks for JEMX. Images, spectra, and light curves were obtained from
these three instruments by using the Integral Science Data Center (ISDC)
official software release (OSA version 3.0). An outline of the software
analysis methods for INTEGRAL data can be found in Goldwurm et al.
(\cite{isgrisoft}), Diehl et al. (\cite{spisoft}) and Wetergaard et al.
(\cite{jemxsoft}).

\subsection{Imaging}

ISGRI detected the source in a total of 280  pointings.  Since each
pointing had an integration time of around 30 minutes, the total ISGRI
observing time was around 500 ks. However, only those detections with a
detection level above 8 were included in the analysis. Thus, the total
time used in our analysis amounts to about 108 ks.

\begin{figure}
   \centering
   \resizebox{\hsize}{!}{\includegraphics[angle=-90]{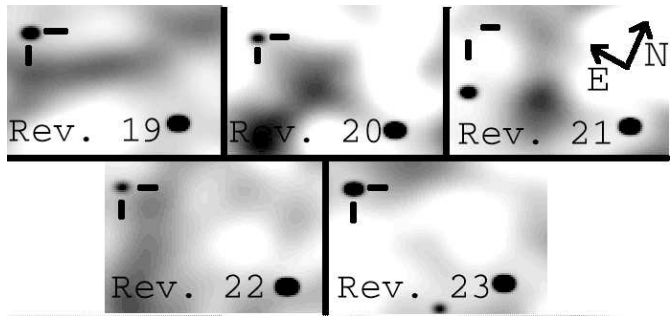}}
      \caption{ Evolution of the detections of \object{SAX J2103.5+4545} with SPI during 
      PV phase, averaging over a revolution. The location of SAX~J2103.5+4545 
      is shown with two orthogonal lines at the top-left corner. The time 
      elapsed between peaks coincides with the orbital period of the system. 
      The source on 
      the right bottom corner is Cyg X-3.}
   \label{spi_evol}
   \end{figure}

Even though the source was inside the SPI field of view during all
December 2002, except during revolution 24 (when the satellite moved into
an empty region to perform background measurements), the averaging of data
over one revolution, gave significant flux values only for revolutions 19,
20, and 23. Figure \ref{spi_evol} depicts the evolution of the
significance of the detection of \object{SAX J2103.5+4545} with SPI. Flux
maxima are modulated with the orbital period of the system.

JEMX only gives positive detections of the source (even forcing the
software to find significant flux at the source position) for 10
pointings, spread over revolutions 23 and 25. This is due to the small field
of view of JEMX and the dithering pattern used.

   \begin{table}
         \label{timaging}
	 \centering
      \caption[]{Most significant  \object{SAX J2103.5+4545} detections.}
    \begin{tabular}{c c c}
    \hline   
Instrument & Mean MJD & Mean Flux \\
  	   &            & phot cm$^{-2}$ s$^{-1}$ keV$^{-1}$ \\
\hline	         
JEMX         & 52630.71 &	 0.00088$\pm$0.00002 \\
(5-20keV)    & 52636.69 &	 0.00021$\pm$0.00001 \\
\hline
ISGRI 	     & 52618.73 &	 0.00037$\pm$0.00004 \\
(20-40 keV)  & 52630.71 &	 0.00025$\pm$0.00002 \\
	     & 52636.69 &	 0.00009$\pm$0.00001 \\
\hline
SPI	     & 52618.73 &	 0.000051$\pm$0.000016 \\
(40-100 keV) & 52621.72 &	 0.000005$\pm$0.000019 \\
	     & 52630.71 &	 0.000067$\pm$0.000010 \\
\hline 
  \end{tabular}
  \end{table}

\begin{table}
         \label{tspectra}
	 \centering
      \caption{Spectra extraction summary. Values for a power-law fitting for all 3 instruments, JEMX, ISGRI and SPI.}
         \begin{tabular}{l @{}c@{} c@{} c @{}c@{} c@{}}
	 \hline
            Instrument & ~Rev. & ~Energy Range & ~Flux ($\times 10^{-10}$)& $\chi ^{2}_{\rm red}$  & $\chi ^{2}_{\rm red}$ \\
                       &            &    (keV)     & erg cm$^{-2}$ s$^{-1}$          & ~ powerlaw & ~bremsstrahlung          \\
            \hline
JEMX	 	&	~23 &	 4-25  &	   5.6	&  0.99 & --\\
\hline
ISGRI		&	~19 &	 25-70  &	   3.5	&  1.3 & 1.2\\
                &  	~23 &	 20-150 &	   5.3	&  1.1 & 1.1 \\
                &  	~25 &	 25-70  &	   1.8	&  1.4 & 1.3 \\
\hline
SPI		&	~19 &	 20-150 &	   7.4	&  1.1 & -- \\
                &  	~23 &	 20-150 &	   8.3	&  1.3  & -- \\
            \hline 

            \end{tabular}
\end{table}

Best ISGRI position is R.A.=21h 03m 31s and DEC=45$^{o}$ 45' 00'', with 
$\sim$1 arcmin error radius.
Best JEMX source location is R.A.=21h 03m 36.7s and DEC=45$^{o}$ 45'
02.7'', with an error radius of 30 arcsec. These values represent a
considerable improvement with respect to the $\sim$ 2 arcmin
uncertainty radius of the BeppoSAX Wide Field Camera.

   \begin{figure}
   \centering
   \resizebox{\hsize}{!}{\includegraphics[angle=-90]{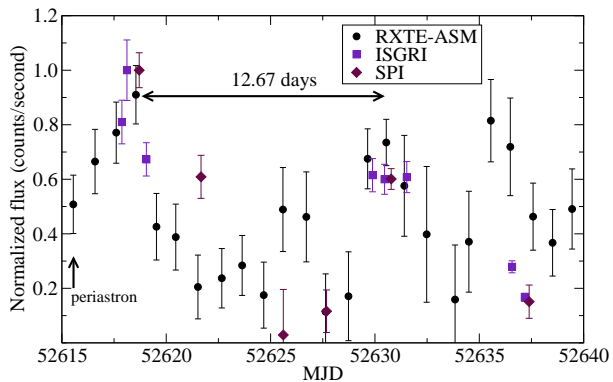}}
      \caption{ISGRI, SPI and ASM RXTE light curve. Each data point represents 
      an averaged one day flux for all the instruments. Units are  
      normalized to the peak value.}
   \label{pvphase_lc}
   \end{figure}

\subsection{Timing}

In revolution 19 (December 8-12, 2002, MJD 52617.39--52620.06), ISGRI data
showed the peak of an outburst. The source was active during previous
revolutions but not detected because it was quite marginal in the field of
view (FOV) and the observing mode was set to {\it staring}, i.e. pointing
continuously to the same direction. The detection of SAX 2103.5+4545 extends to
revolution number 20, although the detection level is too
low. Another outburst peak was detected at the end of revolution 22 and in
revolution number 23 (December 20-23, 2002, MJD 52629.36--52632.05), but
this time the source was well inside the FOV. In about a dozen pointings,
it was inside the fully coded FOV (FCFOV) of ISGRI. During revolution 24 
(December 23-26, 2002, MJD 52632.34-52635.00), the satellite moved again
away from the source, and only at the end of this revolution (December 26, 
2003) and during revolution 25 (December 26-29,  2002, MJD between 52635.37
and 52638.00), the source appears again. Table 1 gives a summary of the
detections for all three high-energy instruments.

The peaks of the two type-I outbursts detected  take place in revolution 19 and 23, i.e., they are 
separated by about 12 days, which corresponds to the value of the orbital
period of the system (Fig.~\ref{spi_evol}). The source lies below the
detection level of the instruments between outbursts. This is the typical
behaviour of a type I outburst seen in most BeX-ray binaries. 

In Fig. \ref{pvphase_lc} one-day averaged fluxes from RXTE-ASM (2-12 keV),
ISGRI (20-40 keV) and SPI (20-40 keV) are shown, with normalized units in
counts per second. INTEGRAL detections coincide with peaks in the
orbital-modulated RXTE-ASM light curve. A careful look at the maxima of
the outbursts shows that the second outburst is weaker than the previous
one.  
Note also the emission that takes place after apastron passage (around MJD
52626 and around MJD 52637).

\subsection{Pulse Period}

\begin{figure}
   \centering
   \resizebox{\hsize}{!}{\includegraphics[angle=-90]{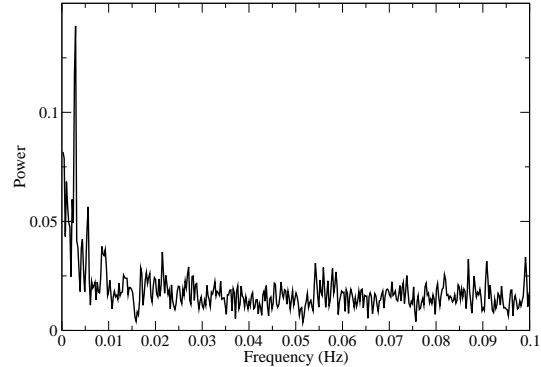}}
      \caption{Power-spectrum for positive detections of \object{SAX J2103.5+4545}
inside the FCFOV of ISGRI during revolution number 23. The maximum
power is achieved at $\sim 0.00281$ Hz.}

   \label{isgri_pulse_period}
   \end{figure}

Pulse period analysis is hindered by the low S/N. ISGRI light curve extraction
in sub-science window time-scales does not give very good results for
off-axis and weak sources. Thus only data from pointings in revolution 23
for which the source was in the FCFOV were used. That makes a total of 17
pointings. An average power spectrum is shown in
Fig.~\ref{isgri_pulse_period}. This power spectrum was obtained by
averaging the Fast Fourier transformed light curves of each pointing.  The
maximum power is found for the frequency 0.00281 Hz, consistent with the
known pulse period of the system.

Once we ensured the presence of pulsations we carried out an epoch folding
analysis of the entire 1-s binned light curve of revolution 23. The
time-span of these observations is almost 2 days. The resulting pulse
period is 354.9$\pm$0.5 seconds, in excellent agreement with that reported
by Inam et al. (2004) of 354.794 seconds. Correction to the solar barycentre was not carried out as this correction is 
much smaller ($\sim 10^{-3}$ s) than the quoted error.

\subsection{Spectral Analysis}

For ISGRI and JEMX, spectra extraction works on a per-pointing basis,
hence all our spectra were extracted for individual pointings and later on
summed up to build averages. In addition, ISGRI can provide reliable
spectra only when the source is inside the fully coded field of view
(FCFOV). For SPI and for a source like \object{SAX J2103.5+4545}, to
achive a S/N of $\sim$10, we need typically about 300 science
windows, namely, a total integration time of around 550 ks. With these
constraints the number of "good" spectra reduces to 22 (17 from ISGRI, 1
from SPI and 4 from JEMX). For SPI up to 2 can be obtained (averaged over
a revolution and for revolutions 19 and 23) but to increase S/N we have
used only the averaged spectrum for all December 2002. A summary of
spectra extraction and flux values for a single power-law fit is presented in Table 2. 

\begin{figure}
\centering
\resizebox{\hsize}{!}{\includegraphics[angle=-90]{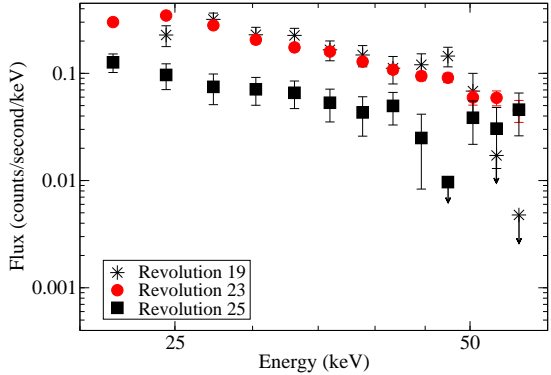}}
\caption{ISGRI \object{SAX J2103.5+4545} spectra in revolutions 19 (stars), 23 (circles) 
and 25 (squares). Apart from a difference in flux no spectral variability
is apparent.}
\label{spe_rev23_fig}
\end{figure}

In order to search for medium-term spectral variability ($\sim$days) we
obtained mean ISGRI spectra for  revolutions 19, 23 and 25 (see Fig.
\ref{spe_rev23_fig} and Table 2). However, no significant changes were
observed, within the errors. These spectra were equally well fitted to a
power law and bremsstrahlung models with photon index $\Gamma$=2.5$\pm$0.1
and $kT=36\pm4$ keV, respectively. A decrease in the 20-50 keV flux from
revolution 23 to revolution 25 is apparent, though. We also searched for
short-term spectral variability ($\sim$hours) by fitting a power-law model
to each pointing of revolution 23 for which we could extract an spectrum (FCFOV observations). The photon spectral index remained
constant, within the errors. A mean 20-100 keV flux of 5.3$\times$10$^{-10}$
erg cm$^2$ s$^{-1}$ was found.  Table \ref{spectra_isgri_r23} gives the
evolution of the power-law index during revolution 23. 

   \begin{table}
      \caption[]{Time evolution of spectral power-law fitted parameters of
      \object{SAX J2103.5+4545} using ISGRI data from revolution 23.}
         \label{spectra_isgri_r23}
	 \centering
         \begin{tabular}{c c  c c}
          \hline
            MJD & $\Gamma$ & $\chi^{2}$ & DOF  \\
         \hline
52629.91931 &	2.4$\pm$ 0.3 &   1.27 &  25 \\
52630.13241 &	2.3$\pm$ 0.3 &   1.24 &  25 \\
52630.15902 &	2.6$\pm$ 0.3 &   0.72 &  25 \\
52630.18565 &	1.9$\pm$ 0.5 &   1.14 &  25 \\
52630.21228 &	1.9$\pm$ 0.3 &   1.70 &  25 \\
52630.37240 &	2.1$\pm$ 0.5 &   1.35 &  25 \\
52630.39909 &	2.2$\pm$ 0.2 &   1.52 &  25 \\
52630.42570 &	2.5$\pm$ 0.3 &   0.96 &  25 \\
52630.45233 &	2.2$\pm$ 0.3 &   0.89 &  25 \\
52631.25349 &	2.4$\pm$ 0.5 &   1.08 &  25 \\
52631.46663 &	2.6$\pm$ 0.4 &   1.03 &  25 \\
52631.49325 &	2.4$\pm$ 0.4 &   0.99 &  25 \\
52631.51988 &	2.3$\pm$ 0.4 &   1.13 &  25 \\
52631.54651 &	2.2$\pm$ 0.3 &   1.37 &  25 \\
52631.79678 &	2.5$\pm$ 0.4 &   1.01 &  25 \\
52631.82341 &	2.6$\pm$ 0.4 &   1.43 &  25 \\
52631.85004 &	2.3$\pm$ 0.4 &   1.30 &  25 \\
         \hline
         \end{tabular}
 
  \end{table}
   
\begin{figure}
\centering
\resizebox{\hsize}{!}{\includegraphics[angle=-90,width=10cm]{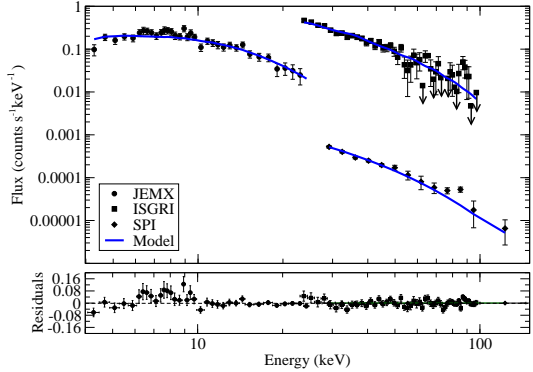}}
\caption{Fit to joint JEMX, ISGRI and SPI spectra. JEMX and ISGRI spectra are mean
spectra for revolution 23. For SPI a mean spectrum for December 2002 is used.}
\label{spe_3inst_fig}
\end{figure}

A 4-150 keV combined spectrum of all three instruments is shown in Fig.
\ref{spe_3inst_fig}, where the mean ISGRI and JEMX spectra of revolution
23 and the mean SPI spectrum of the entire PV phase are shown. The
spectrum was fitted with a power law plus an exponential cut-off, i.e.,
the photon flux density adopts the form,  $f(E)=K E^{-\Gamma} e^{(E_{\rm
cut}-E)/E_{\rm fold}}$,  where $K$ is a normalisation constant. No
absorption was included because previous reported values of the hydrogen
column density, namely ($\sim 3.7 \times 10^{22}$ cm$^{-2}$), do not have
any effect above 4 keV. The resulting best-fit parameters are
$\Gamma=1.0\pm0.1$, $E_{\rm fold}=30.9\pm2.5$ and  $E_{\rm cut}=7.6\pm2.0$
keV, consistent with previous published values (Baykal et al. 2002; Inam
et al. \cite{inam04}). The 4-150 keV X-ray flux, $1.2\times 10^{-9}$ erg
cm$^{-2}$ s$^{-1}$ and the spectral photon index of $1.0\pm 0.1$ indicate
that the source was in a bright state.

There seems to be some excess emission between 6 and 10 keV. The modest
signal-to-noise of the data together with the moderate energy resolution
of the instrument prevent us from identifying it as an iron line feature or
absorption edge. Nevertheless, although the presence of a iron line has
been reported in the past (Baykal et al. \cite{baykal02}) the soft escess
is likely to be an artifact of the reduction process. Sources in the
outer field of view of JEMX are more strongly affected by vigneting
correction and error estimation since they only  illuminate a small
fraction of the detector. This  applies to  \object{SAX J2103.5+4545},
with the closest angular distance to on-axis position being  $\sim$2
degrees. 

   \begin{figure}
   \centering
   \resizebox{\hsize}{!}{\includegraphics[angle=-90]{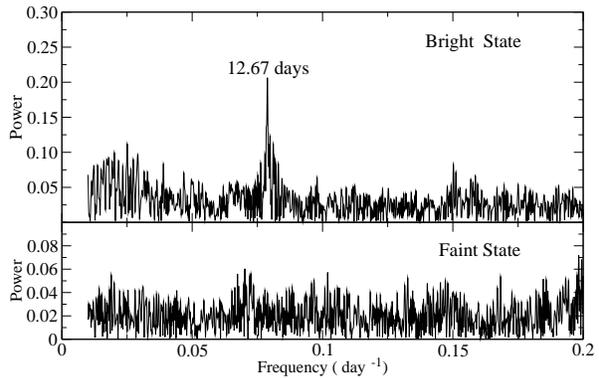}}
      \caption{Comparison of the power spectra of \object{SAX J2103.5+4545}
      during the bright and faint states. Orbital modulated outbursts are only seen 
      during bright states.}
   \label{long}
   \end{figure}

\section{Discussion}

We have carried out a timing and spectral analysis of the two outbursts
that the Be/X-ray binary \object{SAX J2103.5+4545} underwent during the
INTEGRAL Performance and Verification phase in December 2002. The
excellent imaging capabilities of INTEGRAL allowed us to reduce the 
uncertainty in the position of the source by a factor of $\sim 3$. The
INTEGRAL location of the X/$\gamma$-ray source agree with the position of
the recently proposed optical counterpart (Reig et al. \cite{reig04}).

No spectral variability on time scales of hours/days has been observed in
the INTEGRAL data of \object{SAX J2103.5+4545}. In contrast, the source is
quite rich in temporal variability. In the short term ($\sim$seconds),
X-ray pulsations with a pulse period of 354.9 s are clearly detected. This
value of the pulse period contrasts with that measured by Baykal et al.
(2002) of 358.6 s and indicates a long-term spin-up of the neutron star.
In this respect, it is illustrative to compare our results with the quasi
simultaneous RXTE and XMM observations (and contemporaneous to the
INTEGRAL observations)  of Inam et al. (2004). They found a pulse period
of 354.8 s, in excellent agreement with the one we obtained. Although a
short spin-down phase was found by Baykal et al. (2002), it seems that the
source has been continuously accelerating by the in-falling accretion of
matter. The change in pulse period would be compatible with a continuous
spin-up of 2.5$\times$10$^{-13}$ Hz s$^{-1}$ since 1999. Nevertheless, the
spin-up is known to be linked to the accretion process, and hence, it is
discontinuous and depends on the dynamics of the system, including both
the neutron star orbit and the Be star dense envelope. Unlike most BeX
systems (Corbet 1986), \object{SAX J2103.5+4545} does not rotate at the
equilibrium period (Reig et al. \cite{reig04}). 

In the medium term ($\sim$days), the X-ray behaviour of \object{SAX
J2103.5+4545} is characterised by regular increases of the X-ray flux
modulated with the orbital period (type I outbursts). The INTEGRAL
observations coincided with two of these outbursts. Assuming a distance to
the source of 6.5$\pm$0.9 kpc, (Reig et al. \cite{reig04}) the X-ray luminosity in
the 4-150 keV energy range, amounts to $\sim$6.0 $\times$ 10$^{36}$ erg
s$^{-1}$, which is typical of type I outbursts in BeX systems. The quoted error in the distance introduces a 
30\% of relative error in the luminosity determination.

\object{SAX J2103.5+4545} also displays longer-term X-ray variability
($\sim$months), consisting of low and high-activity X-ray states. Type I
outbursts are only seen during bright states (Fig.~\ref{long}, see also
Baykal et al. \cite{baykal02}). Given the short orbital period (12.7 days)
the neutron star must exert  substantial influence on the evolution of the
circumstellar disk. In the optical band, this influence translates into a
highly variable H$\alpha$ line, exhibiting V/R asymmetry and reversing
from emission into absorption on time scales of a few days (Reig et al.
2004).  In the X-ray band, the influence of the neutron star on the Be
star envelope might be at the origin of the lack of giant (type II)
outbursts and the long-term activity. In the framework of the viscous
decretion model (Okazaki \& Negueruela \cite{oka01}) the scenario would
then be as follows. The tidal interaction of the neutron star produces the
truncation of the Be star envelope. The truncation radius in \object{SAX
J2103.5+4545} would be similar in size to the critical lobe radius at
periastron. When the density and/or size of the Be star envelope are large
enough, matter fills the critical lobe and is accreted onto the neutron
star.  The system is in the bright state. One periastron passage is not
enough to exhaust the fuel from the disk. However,  after several orbits
the Be star's disk weakens, the amount of matter available for accretion
decreases and the system enters a faint state. Once the disk recovers the
initial conditions the cycle starts again.

\begin{figure}
   \centering
   \resizebox{\hsize}{!}{\includegraphics[angle=-90]{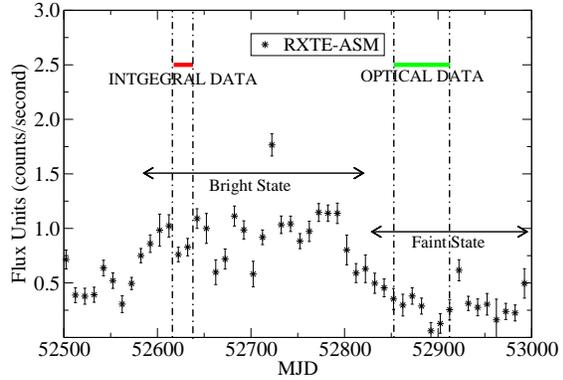}}
      \caption{10 day averaged RXTE-ASM light curve showing the transition 
      from bright state (end of 2002 and first half of 2003) into faint 
      state (second half of 2003). The epochs of INTEGRAL and optical observations 
      are indicated by horizontal lines.}
   \label{asm_pablo}
   \end{figure}

Although the number of optical observations of \object{SAX J2103.5+4545}
is scarce, the information available indicates a correlation between the
long-term optical and X/$\gamma$-ray variability. The optical 
observations (August and September 2003) reported by Reig et al. (2004)
coincided with the transition to an X-ray faint state (see
Fig.~\ref{asm_pablo}). The H$\alpha$ emission was low (equivalent width of
$\sim -1.5$), even the H$\alpha$ line appeared in absorption and the
optical magnitudes were consistent with little circumstellar reddening. In
contrast, the photon spectral index and the X-ray flux obtained from the
spectral analysis of the INTEGRAL observations indicate that \object{SAX
J2103.5+4545} was in the bright state.  Unfortunately, by the time of the
INTEGRAL observations, the optical counterpart to \object{SAX J2103.5+4545} was not included in the
Optical Monitoring Camera catalogue and no
optical data are available.

\section{Conclusion}

The BeX-ray binary \object{SAX J2103.5+4545} has been observed with
INTEGRAL. Our main results are: {\it i)} we have reduced to about 30 arc
seconds the uncertainty in the position of the system. This new position
agrees with the newly discovered optical counterpart, {\it ii)} we have
obtained the first broad-band spectrum (4-150 keV) of \object{SAX
J2103.5+4545}. Above 150 keV the flux is comparable to the sensitivity
limits of the instruments. Further improvements in the software and/or in
the response matrices, or longer exposure times, may allow to reach higher
energies, {\it iii)} the value of the pulse period indicates a long-term
spin-up episode and {\it iv)} the high-energy emission correlates with the
optical activity state of the system.

Simultaneous observations in the optical/IR and X/$\gamma$-ray bands are
needed to understand the connection between the changes in the Be star
envelope and the behavior in the high-energy bands. In particular,
simultaneous optical/X-ray observations of \object{SAX J2103.5+4545}
during the bright state would be very clarifying. This system, with
the shortest orbital period known for an accretion-powered BeX binary, represents
an excellent laboratory to test the current models for BeX systems.

\begin{acknowledgements}
      Part of this work was supported by project number 
ESP-2002-04124-C03-02 "INTEGRAL. OPERACIONES C1". P.R. acknowledges partial
support from
the programme {\it Ram\'on y Cajal} funded by the Spanish Ministery of
Science and Technology and the University of Valencia, under grant
ESP2002-04124-C03-01. 
\end{acknowledgements}


\begin{thebibliography}{}
      
   \bibitem[2000]{baykal00} Baykal, A., Stark, M.J. and Swank, J., 2000,
      ApJ, 544, L129

   \bibitem[2002]{baykal02}  Baykal, A., Stark ,M.J. and Swank, J.,2002,
     ApJ, 569, 903

  \bibitem[1986]{corbet96} Corbet, R.D.  1986, 
      MNRAS, 220, 1047 

   \bibitem[2003]{santo03} Del Santo, M., Rodriguez, J., Ubertini, P. et al, 2003,
      A\&A, 411, L339

   \bibitem[2003]{spisoft} Diehl, R., Baby, N., Beckmann, V.  et al, 2003,
      A\&A, 411, L117   
   
   \bibitem[2003]{isgrisoft} Goldwurm, A., David,, P., Foschini, L. et al, 2003,
      A\&A, 411, L223
      
   \bibitem[2003]{gros03} Gros, A., Goldwurm, A., Cadolle-Bel, M., Goldoni, P., Rodriguez, J., Foschini, L., Del Santo, M., Blay, P., 2003,
      A\&A, 411, L179    

  \bibitem[1998]{hulleman98} Hulleman, F., in 't Zand, J.J.M. and Heise, J., 1998,
      A\&A, 337, L25
 
   \bibitem[2004]{inam04} Inam, S.C., Baykal, A., Swank, J., and Stark, M.J. 2004,
      astro-ph/0402221

   \bibitem[2003]{jemxpaper} Lund, N., Budtz-Jørgensen, C., Westergaard, N. J. et al, 2003,
      A\&A, 411, L231
      
   \bibitem[2003]{luto03} Lutovinov, A., Molkov, S. and Revnistev, M., 2003,
      AstL, 29, 713

    \bibitem[2003]{omcpaper} Mas-Hesse, J.M.,  Giménez, A., Culhane, J. L. et al, 2003
      A\&A, 411, L261
      
    \bibitem[2001]{oka01} Okazaki, A. T. and Negueruela, I.,  2001, A\&A,
    377, 161

   \bibitem[2003]{reig03} Reig P. and Mavromatakis F., 2003, 
     ATEL, 173 

   \bibitem[2004]{reig04} Reig, P., Negueruela, I., Fabregat, J., Chato, R., Blay, P. and Mavromatakis, F., 2004,
     A\&A, in press.     
   
    \bibitem[2003]{spiros} Skinner, G. and Connell, P., 2003,
      A\&A, 411, L123

   \bibitem[2003]{ibispaper} Ubertini, P., Lebrun ,F., Di Cocco, G. et al,  2003,
      A\&A, 411, L131
   
   \bibitem[2003]{spipaper} Vedrenne, G., Roques, J.-P., Schönfelder, V. et al, 2003,
      A\&A, 411, L63

   \bibitem[2003]{jemxsoft} Westergaard, N.J., Kretschmar, P., Oxborrow, C. A. et al, 2003,
      A\&A, 411, L257

    \bibitem[2003]{integral} Winkler, C., Courvoisier, T. J.-L., Di Cocco, G. et al, 2003,
      A\&A, 411, L1

\end{thebibliography}
\end{document}